\documentclass[sigconf,authorversion]{acmart}

\AtBeginDocument{%
  \providecommand\BibTeX{{%
    \normalfont B\kern-0.5em{\scshape i\kern-0.25em b}\kern-0.8em\TeX}}}

\copyrightyear{2021} 
\acmYear{2021} 
\setcopyright{acmlicensed}\acmConference[CHI '21]{CHI Conference on Human Factors in Computing Systems}{May 8--13, 2021}{Yokohama, Japan}
\acmBooktitle{CHI Conference on Human Factors in Computing Systems (CHI '21), May 8--13, 2021, Yokohama, Japan}
\acmPrice{15.00}
\acmDOI{10.1145/3411764.3445673}
\acmISBN{978-1-4503-8096-6/21/05}

\newcommand{\scale}{Technology-Supported Reflection Inventory}
\newcommand{\scaleshort}{TSRI}

\begin{document}

\title{The Development and Validation of the Technology-Supported Reflection Inventory}


\author{Marit Bentvelzen}
\affiliation{%
    \institution{Utrecht University}
    \city{Utrecht}
    \country{the Netherlands}}
\email{m.bentvelzen@uu.nl}

\author{Jasmin Niess}
\affiliation{%
    \institution{University of Bremen}
    \city{Bremen}
    \country{Germany}}
\email{niessj@uni-bremen.de}

\author{Miko\l{}aj P. Wo\'{z}niak}
\affiliation{%
    \institution{Lodz University of Technology}
    \city{Lodz}
    \country{Poland}}
\email{mpwozniak@ubicomp.pl}

\author{Pawe\l{} W. Wo\'{z}niak}
\affiliation{%
    \institution{Utrecht University}
    \city{Utrecht}
    \country{the Netherlands}}
\email{p.w.wozniak@uu.nl}

\begin{abstract}
Reflection is an often addressed design goal in Human-Computer Interaction (HCI) research. An increasing number of artefacts for reflection have been developed in recent years. However, evaluating if and how an interactive technology helps a user reflect is still complex. This makes it difficult to compare artefacts (or prototypes) for reflection, impeding future design efforts. To address this issue, we developed the \emph{Technology-Supported Reflection Inventory} (TSRI), which is a scale that evaluates how effectively a system supports reflection. We first created a list of possible scale items based on past work in defining reflection. The items were then reviewed by experts. Next, we performed exploratory factor analysis to reduce the scale to its final length of nine items. Subsequently, we confirmed test-retest validity of our instrument, as well as its construct validity. The TSRI enables researchers and practitioners to compare prototypes designed to support reflection.  
\end{abstract}


\maketitle

\section{Introduction}
As an increasing number of systems around us aim to increase our well-being, reflection is a concept that gains more and more relevance. While reflection is complex and can be interpreted in a variety of ways, it is regarded as beneficial and desired~\cite{Baumer_reflectiveinformatics}. As a consequence, recent years have seen an ever-increasing interest in HCI to design technology that supports reflection in users. These design efforts span a wide spectrum of applications, including mental health~\cite{Thieme2012weveLifestyles}, personal informatics~\cite{Li_2010_new}, health~\cite{Baumer2012PrescriptiveHealth, Ayobi2020TracklySelf-Care}, as well as reflection on daily life~\cite{Mols2020EverydayDott}. There is an emerging challenge in the field to understand what particular design qualities can foster reflection.

However, the evaluation of reflection support technologies is often limited. Consequently, it is often unclear how a certain design affected the user's ability to reflect. A systematic review by Baumer et al.~\cite{Baumer2014ReviewingDesign} showed that there was no consensus in the field regarding the evaluation of reflection support technology. Studies that used quantitative assessments rarely measured reflection, but rather focused on user experience. Qualitative evaluation is an often-used alternative. Yet, this approach, encounters similar difficulties, namely, the lack of a clear definition of reflection and not asking users directly about reflection. The lack of proper evaluation methods for reflection support technology makes it difficult to compare different design alternatives. This, in turn, complicates future design efforts. Consequently, there is a need for building rapid and standardised means of evaluating whether a system can support reflection.

To this end, this paper reports on the \emph{Technology-Supported Reflection Inventory} (TSRI), a scale that evaluates how an interactive system fosters personal, data-driven reflection. We first investigated past work in reflection theory, existing reflection scales and studies regarding reflection in personal informatics. Subsequently, we generated initial items for the scale, which were then subjected to two rounds of expert reviews. We then performed exploratory factor analysis in order to reduce the number of items and obtain the final scale. Finally, we evaluated the TSRI by testing its reliability and construct validity. Our work offers the first, to our knowledge, validated system-centric scale for evaluation of HCI technologies designed for reflection. We contribute the scale for its future use by the research and practitioner community along with a systematic validation of the instrument.

This work is organised as follows, we begin by discussing related work associated with designing for reflection in HCI and existing scales for reflection. Next, we describe the process of developing the scale and the use of an exploratory factor analysis to reduce and optimise the number of scale items. Subsequently, we report on the validation of the scale, and, finally, discuss on the use and limitations of the \scaleshort{}. 

\section{Related Work}
In this section, we first provide an overview of the general understanding of reflection within HCI. We then review past work in the area of personal informatics and showcase
that there is a need for instruments that enable researches to evaluate technologies designed for reflection.

\subsection{Understanding Reflection}
Reflection is a recurring research theme in HCI research. Yet, as a systematic review by Baumer et al.~\cite{Baumer2014ReviewingDesign} showed, few papers (30 out of 76) clearly define the concept of reflection. Consequently, within the HCI community, there are multiple understandings of reflection in place~\cite{Baumer_reflectiveinformatics}. 
Several conceptual and theoretical accounts of reflection co-exist in HCI~\cite{Slovak2017ReflectiveReflection, Baumer2014ReviewingDesign}.

The majority of HCI papers employ Schön’s~\cite{Schon1983ThePractitioner} notion of reflection-in-action and reflection-on-action. However, there is no consensus on the definition of reflection~\cite{Baumer2014ReviewingDesign}, and past research also used alternative definitions by Dewey~\cite{Clayphan2017ASpace, Baumer_reflectiveinformatics}, Moon~\cite{Fleck2010ReflectingLandscape}, Mezirow~\cite{Mols2016InformingPractices, Ortiz2018EnablingChimeria:Grayscale, Webb_2013_new} and Boud~\cite{Rivera-Pelayo2017IntroducingWork}. These four definitions regard reflection as a process that happens in hindsight, after the experience has taken place. These definitions are more in line with reflection-on-action, as opposed to Schön's notion of reflection-in-action. 
This multitude of approaches used shows how the field desires to conceptualise reflection in order to understand how to design technologies that could support users in reflecting. This work contributes to building a better understanding of reflection in the context of interactive technologies by building a scale that identifies reflection-supporting qualities in designs.

\subsection{Designing for Reflection}
Reflection is also a concept in past studies of of personal informatics experiences. 
Li's \emph{Stage-Based Model of Personal Informatics Systems}~\cite{Li_2010_new} considers reflection to be a stage that occurs after the \emph{preparation}, \emph{collection} and \emph{integration} of personal data. Subsequently, after reflecting on personal data a user then moves on to \emph{action}, the final stage of the stage-based model~\cite{Li_2010_new}. Li's model was later extended by Epstein et al., who proposed the \emph{Lived Informatics Model of Personal Informatics}~\cite{Epstein2015AInformatics}. This model includes additional stages, such as deciding to track, the selection of tools, and defines \emph{tracking and acting} as an ongoing process of collection, integration and tracking~\cite{Epstein2015AInformatics}. 
These two models offer a high-level overview, but do not provide an in-depth understanding of what the reflection stage entails. As previously noted by Baumer~\cite{Baumer2014ReviewingDesign}, these models carry an implicit assumption that reflection automatically occurs as long as a user has access to data that has been `prepared, combined, and transformed'~\cite{Baumer2014ReviewingDesign}. However, this conflicts with reflection theories, that highlight that reflection often does not occur automatically, but needs to be encouraged, as observed by Slovak~\cite{Slovak2017ReflectiveReflection}. This is also in line with Niess and Wo\'{z}niak~\cite{Niess2018SupportingModel}, who concluded that reflection with technology support was instrumental for achieving fitness tacker goals.

Concurrently, the HCI field contributed a number of systems that were reported to effectively facilitate reflection, e.g. ~\cite{Isaacs2013EchoesWell-being, Baumer2012PrescriptiveHealth, Ayobi2020TracklySelf-Care, Arakawa2020INWARD:Coaching}.
An example of such a system is \emph{Trackly}, designed by Ayobi et al.~\cite{Ayobi2020TracklySelf-Care}. Trackly is a mobile app that helps patients with Multiple Sclerosis to manually track their symptoms and to reflect on the collected data through visualisations. Whereas Trackly is mostly used in an individualistic setting, there are also systems that use social interaction to enhance reflection, such as \emph{MoodMap}, designed by Rivera-Pelayo et al.~\cite{Rivera-Pelayo2017IntroducingWork}. MoodMap is a computer application that lets users track and reflect on their mood at work. In this sense, MoodMap is comparable to Trackly, yet, this system differs in that it also lets users compare their mood to that of colleagues. Another example is \emph{Reveal-it!} designed by Valkanova et al.~\cite{Valkanova_2013_new}, which is an interactive public display that encourages participants to reflect on their energy consumption to increase social awareness and discourse. The system lets a user voluntarily enter their energy consumption data, which is then visualised as a sunburst representation on the public display. This makes it possible to compare one’s energy consumption with others as well as more general statistics for neighbourhoods, which in turn can lead to reflection.

The examples above suggests that there are certain qualities to interactive technologies that determine how effectively they can support reflection. Reflection systems address different application areas and use different interaction techniques. Despite this diversity, reflection appears to a unifying quality. Further, the design process of such technologies would benefit from comparing prototypes at different stages of the design to determine optimal alternatives. However, as Baumer~\cite{Baumer2014ReviewingDesign} remarked, there are no established ways of evaluating systems that support reflection. Our work aims to contribute to a better understanding of reflection by building a validated evaluation method for systems that support reflection.

\subsection{Measuring Reflection}
As we aim to develop a measure of reflection useful for HCI work, we first examine measures of reflection in other fields. Several scales were developed with the intention to measure a persons reflective capacity. The Self-Reflection and Insight Scale (SRIS)~\cite{Grant2002TheSelf-consciousness} consists of three factors: \emph{engagement in self-reflection}, \emph{need for reflection} and \emph{insight}. The scale was developed to provide researchers with an instrument to better understand of the socio-cognitive and meta-cognitive processes central to individual change~\cite{Grant2002TheSelf-consciousness}. 

Further, the Groningen Reflection Ability Scale (GRAS)~\cite{Aukes2007TheEducation} was developed to measure the personal reflection ability of medical students. The GRAS consists of 23 items, such as: \emph{'I can see an experience from different standpoints'} and \emph{'I take a closer look at my own habits of thinking'}. The GRAS measures a persons ability to reflect on three dimensions: self-reflection, empathetic reflection and reflective communication. Self-reflection is the introspective aspect of reflection, in which a person frames one's feelings, thoughts, beliefs and norms. In contrast, empathetic reflection is the social extension of self-reflection, the kind of reflection in which a person considers the position of others. Additionally, the GRAS measures reflective communication, which is the behavioural expression of both self-reflection and empathetic reflection~\cite{Aukes2007TheEducation}. 

The SRIS and GRAS both measure a person's reflective capacity. They are based on insight from Psychology that determined that the likelihood of reflecting is tied to one's personality. However, the examples of systems that we reviewed before show that interactive technologies can amplify or enable a user's capacity for reflection. Thus, the related work suggests that there are certain qualities that make interactive technologies more or less effective in supporting reflection. However, these scales were developed with participants of specific professional background in medicine and psychology, consequently being best suited for use in these contexts. 
Consequently, there is a need for a \emph{'artefact-centric'} instrument that enables the evaluation of technology-supported reflection and measures the qualities of an interactive artefact that lead to reflection. A technology-oriented tool, created in a structured process and validated using various methods (eg. confirmatory factor analysis, test-retest procedure and others) may enable more capable and adequate analysis of reflection in HCI systems, especially within personal informatics field.

\section{Method}
Through exploring the literature we identified concepts related to reflection, which we, in turn, used for generating scale items. A starting point for this inquiry was a study by Li et al.~\cite{Li_2010_new} who conceptualised reflection in personal informatics. Their work presents barriers to reflection that should be addressed in the design of interactive systems, including \emph{lack of time} and \emph{effective visualisation of personal data}. Yet, as Baumer et al.~\cite{Baumer2014ReviewingDesign} noted, there is much to be gained from being explicit in what we mean by reflection. Instead of seeing reflection as a natural consequence or by-product of presenting information to users, he argues that designers of interactive systems should engage with reflection as an important part of a larger set of processes and practices. Earlier work by Baumer~\cite{Baumer_reflectiveinformatics} further disentangles reflection into three dimensions: breakdown, inquiry and transformation. Breakdown refers to situations of doubt or puzzlement that grab people’s attention, and can lead to moments of reflection. Inquiry refers to the process of conscious, intentional inquiry of past experiences in which a person reviews a certain situation. Lastly, transformation exemplifies a change in fundamental assumptions and behaviour. Baumer's stance was in line with earlier work by Fleck and Fitzpatrick~\cite{Fleck2010ReflectingLandscape}, who mapped reflection into levels of reflection and conditions for reflection. The levels of reflection consist of five consecutive levels that describe a spectrum of reflective thought, ranging from no reflection to critical reflection. Next to the levels, they also mentioned that creating the right environment for reflection was important, resulting in three conditions for reflection: time, development and encouragement~\cite{Fleck2010ReflectingLandscape}. 

Next to these dimensions, levels and conditions for reflection, we also used reflection literature from Schön~\cite{Schon1983ThePractitioner} as well as existing scales for reflection~\cite{Aukes2007TheEducation, Grant2002TheSelf-consciousness} as inspiration for the initial set of items.

\section{Scale Formation}
This section describes the process of building the \scale{} from conceptual considerations to a final list of items.

\begin{figure}
    \centering
    \includegraphics[scale = 0.50]{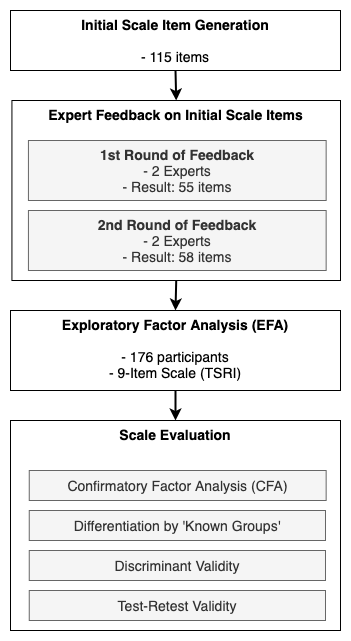}
    \Description{Flowchart depicts the scale formation process that was used. Consists of four rectangles, connected with arrows from top to bottom. Each rectangle represents a phase.}
    \caption{The scale formation process consisted of four phases: (1) the initial generation of scale items, (2) two rounds of expert feedback, (3) Exploratory Factor Analysis, and (4) the evaluation of the scale.}
    \label{fig:flow}
\end{figure}

\subsection{Generating Initial Items}
Three researchers, each experienced in empirical research in technology for reflection from a variety of academic backgrounds, participated in generating initial items for the scale, which is in line with Boateng et al.~\cite{Boateng2018BestPF}. Each researcher first created items, based on related work, on their own. We then arranged a discussion session in which all the generated items were merged and discussed. Through these discussions we removed duplicates. In integrative discussions, we obtained an initial list of 115 items. 

\subsection{Expert Review}
Subsequently, we conducted two rounds of expert reviews to receive feedback on the list of possible scale items. We chose a diverse set of experts to collect broad feedback on the items. The experts in the first round of feedback were a professor in computer science and a researcher in HCI. They provided feedback by commenting on the list of items and suggesting new items. After collecting the feedback from this first expert review we discussed the list of items, identified items that were problematic and either altered or removed these. This process resulted in a list of 55 items. We then organised a second round of feedback, that followed the same process, yet this time with two researchers in psychology. The feedback of this second expert review was then used to finalise the list of scale items, resulting in a final list of 58 items. 

\subsection{Survey}
After the two expert reviews, we designed an online survey using the Qualtrics XM platform to gather data from participants. We then used this data to perform exploratory factor analysis and item reduction. Since there is a lacking consensus over expected local sample size standards, we aimed for a sample size of 100--200 participants, in line with the development and validation of other scales in HCI~\cite{Mejia_Yarosh_2017, marsden_2013, Tondello_2016}.

\begin{table*}
\caption{The \scaleshort{} is a nine-item instrument consisting of three dimensions}
\centering
\begin{tabular}{ll}
\toprule
\textbf{Subscale/Item    }                                                     & \textbf{Factor Loading} \\
\midrule
\textbf{Insight, $\alpha = 0.74$    }                                   &                \\
\midrule
Q1: Using the system has led to a wake-up call to make changes in my life & 0.62\\
Q2: As a result of using the system, I have changed how I approach things & 0.69\\
Q3: Using the system gives me ideas on how to overcome challenges & 0.65 \\
\midrule

\textbf{Exploration, $\alpha = 0.72$    }                                   &                \\
\midrule
Q4: I enjoy exploring my data with the system & 0.68\\
Q5: The system makes it easy to get an overview of my personal data & 0.68\\
Q6: The system makes it easy to review my long-term personal data & 0.65\\
\midrule

\textbf{Comparison, $\alpha = 0.73$    }                                 &                \\
\midrule
Q7: I reflect on my data in the system with others & 0.65\\
Q8: The system helps me to discuss my data with others & 0.64\\
Q9: The system makes me think about how my personal data relates with that of others & 0.71\\

\bottomrule
\end{tabular}
\Description{Table shows in the first column the three subscales, being Insight, Exploration and Comparison. Each subscale consists of three scale-items. The second column shows the factor loading of each scale-item.}
\label{tab:scale}
\end{table*}

\subsubsection{Participants}
We recruited a total number of $n=176$ participants. The participants were recruited using the Amazon Mechanical Turk Service (MTurk) and reimbursed with USD 1. Out of these participants, 93 resided in the USA, 44 in Europe, 38 in Canada and 1 in India. All participants were informed that participation in the study was voluntary and that they could abort the study at any point. We also informed participants that the data would be collected in anonymised form. The survey was conducted online an could be completed in 10 minutes. The average age of participants was $M=34.1$ ($SD = 10.2$) with 62 of participants identifying as female, 113 as male and 1 participant as non-binary. We recruited a sample size within the expected range, exceeding the ones used for development of related scales in recent studies in HCI~\cite{Mejia_Yarosh_2017, Yarosh2014CSCW, Suh2016Developing}. Moreover, our recruitment approach enabled us to reach diverse participant group of various cultural and professional backgrounds, providing more holistic insight than for samples based on student cohorts~\cite{Grant2002TheSelf-consciousness, Aukes2007TheEducation}. 

\subsubsection{Apparatus}
We used the Qualtrics platform to design the online survey, the full survey is available in the auxiliary material. 
To ensure that participants used some form of an interactive technology that can facilitate reflection, we indicated on MTurk that participants needed to use tracking technology (i.e. smartwatch, tracking app) to partake in the survey~\footnote{An overview of the tracking technologies used by our participant is added to the supplementary materials}. The survey consisted of the 55 scale items. Participants scored each item on a 7-point Likert scale, ranging from strongly disagree to strongly agree. 
Participants were advised to complete the online survey on a computer or tablet, but could also use their smartphone. 

\subsection{Exploratory Factor Analysis}
We conducted factor analysis using a varimax rotation, in which we replicated the approach by Mejia and Yarosh~\cite{Mejia_Yarosh_2017}. We then used visual inspection of screen plots to determine the optimal number of factors. After that, we reduced the number of items. We began by removing loadings below $0.40$~\cite{Boateng2018BestPF} and items that loaded on multiple factors. 
We then used an iterative process of removing low loading items and optimising for inter-item reliability in order to further refine the list of items. Current and theoretical Cronbach's alpha coefficients were computed. Our aim was to make the scale as short as possible while maintaining internal consistency and factor structure. This resulted in a loading of three items per factor, and a Cronbach's alpha of $\alpha = 0.78$ for the scale. The theoretical factor model hadcorrect parameters at $TLI=0.97$ and $RMSEA = 0.04$, cf~\cite{Boateng2018BestPF}. Table~\ref{tab:scale} shows the composition of the scale. We determined the names of the scale dimensions in an iterative discussion with three researchers, mapping them to concepts from our literature review.

\section{The TSRI}
The final scale consists of three dimensions that describe the qualities of an interactive technology which supports reflection. Table~\ref{tab:scale} shows the items included in the theoretical final scale.

\subsubsection{Insight}
This dimension describes to what extent the interactive technology offers insight to users. It evaluates if the system provides users with information that is insightful to them, and whether it inspires them to look at situations from a new perspective or makes changes to their lives.

\subsubsection{Exploration}
The second dimension of the \scaleshort{} is exploration. It evaluates the ease and enjoyment of exploring personal data in the system. In order for users to reflect on their personal data, it should be easy for users to access and review their data. This dimension can therefore be seen as a necessary condition for reflection on personal data, which echoes findings by Fleck and Fitzpatrick~\cite{Fleck2010ReflectingLandscape}.

\subsubsection{Comparison}
The third dimension of the \scaleshort{} measures the social dimension of a refection-supporting interactive system, i.e. the extent to which users can compare themselves to others. It is often argued that current interactive technologies rarely support reflection as a social activity, but instead focus on individual cognitive processes~\cite{Mols2016InformingPractices, Baumer2014ReviewingDesign}. Yet, reflection can be encouraged by conversations with others~\cite{book, Mols2016InformingPractices, Fleck2010ReflectingLandscape}. Conversations can provide reflectors with an additional or alternative perspective~\cite{Fleck2010ReflectingLandscape}. This was also shown in a study by Mols~\cite{Mols2016InformingPractices}, $85\%$ of their participants stated that they reflect through conversations with for instance a partner, friend or colleague.

\begin{figure*}
    \centering
    \includegraphics[width=.9\textwidth]{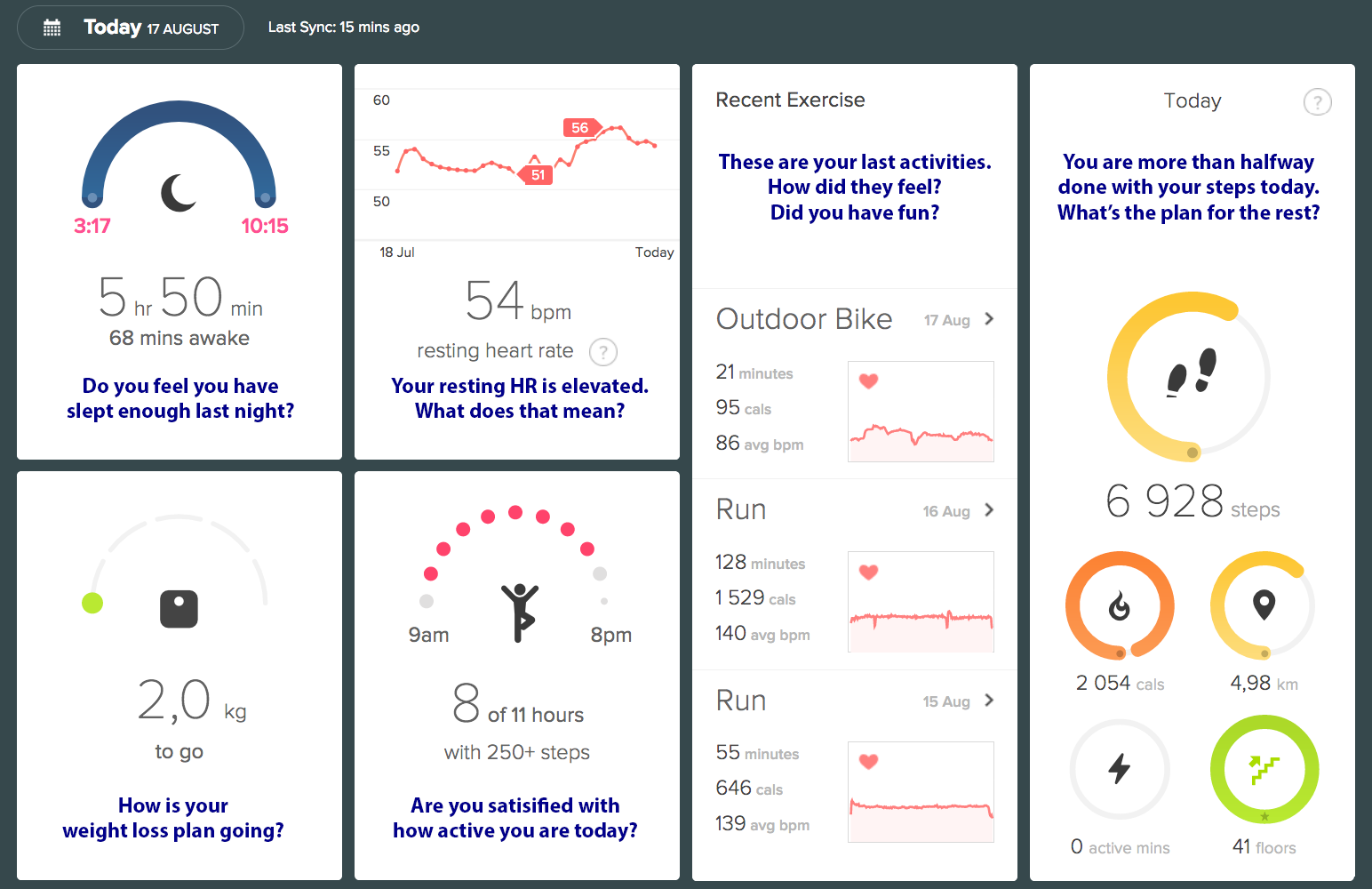}
    \Description{Dashboard that visualises the user's personal data. It visualizes sleep performance data, resting heart rate, recent exercise, step data, weight loss goal and the number of hours with more than 250 steps. For each category text has been added to enhance reflection on the visualisations.}
    \caption{The modified dashboard which we used for testing \scaleshort{}'s construct validity. The modified dashboard is the standard Fitbit dashboard enriched with questions (which is a known technique for enhancing reflection~\cite{book, Mols2016InformingPractices}) regarding the personal data depicted by the dashboard.}
    \label{fig:modified_dashboard}
\end{figure*}

\section{Scale Validity}
After building the theoretical composition of the scale, we continued with the evaluation of the \scaleshort. We conducted Confirmatory Factor Analysis to verify the underlying model. This was followed by a series of tests to check the scale's construct validity and reliability.

\subsection{Participants}
We recruited $n=56$ participants via MTurk, using the same approach as in the first survey. The study was conducted online. The reimbursement for completing the survey was USD 0.80. The average age of the participants was $M=36.6$, $SD = 11.8$, 23 participants identified as female and 33 as male.

\subsection{Apparatus}
To evaluate the scale, we created a dashboard that showed an overview of personal data from a fitness tracker. We developed two versions; a standard dashboard, and a modified dashboard (see Figure~\ref{fig:modified_dashboard}). The modified dashboard included questions to enhance reflection. We chose to enrich the dashboard with questions, because this is a known technique for reflection~\cite{book, Mols2016InformingPractices}. We assumed the modified dashboard would encourage users more to reflect on their data when compared to the standard dashboard. We randomly presented each participant with one of the two dashboards and a short description: `Below, you see some information produced by Brian's fitness tracker. For the rest of the survey, we ask you to become Brian.' Afterwards, we asked participants to `put themselves in the role of Brian' and to indicate how much they agreed with each item of the (then hypothetical) \scaleshort{} about the presented dashboard on a 7-point Likert scale (strongly agree to strongly disagree).

\subsection{Confirmatory Factor Analysis}
To investigate the correctness of the \scaleshort{} factor model, we conducted Confirmatory Factor Analysis (CFA). The resulting Tucker Lewis Index of factoring reliability was $TLI=0.91$, which is an acceptable fit~\cite{Boateng2018BestPF}. We believe that recruiting additional participants and conducting the CFA enabled us to reinforce the reliability of \scaleshort{}, even though we recognise this is not a standard approach in developing scales for HCI (eg.~\cite{Mejia_Yarosh_2017,haynes1995content,Suh2016Developing} did not apply this method).

\subsection{Differentiation by `Known Groups'}
Next, we used `known groups'~\cite{Boateng2018BestPF} to contribute to establishing the construct validity of \scaleshort{}. To that end, we conducted t-tests to investigate if the TSRI scores differed between the two designs of the tracker dashboard. We found that the modified dashboard scored higher on the full scale as well as on the Insight and Exploration sub-scales. Table~\ref{tab:groups} shows the results.

\begin{table*}
\centering
\caption{Construct validity assessment differentiation by known groups with t-tests. Bonferroni-corrected p-values are reported.}
\begin{tabular}{lllllll}
\toprule
Scale/Sub-scale & $M_{Modified}$ & $SD_{Modified}$ &  $M_{Standard}$ & $SD_{Standard}$ & $t(54)$      & $p$      \\
\midrule
TSRI        & 46.89 & 8.86          & 42.00 & 8.04     & $2.16$ & $0.041$ \\
TSRI-Insight        & 17.14 & 3.04         & 15.04&  2.97      & $2.62$ & $0.020$  \\
TSRI-Exploration        & 14.86  &3.80            & 12.75&  3.69     & $2.11$ & $0.028$ \\
TSRI-Comparison        & 14.89  & 4.18            & 14.21 & 3.34     & $0.67$ & $0.477$\\
\bottomrule
\end{tabular}
\Description{Table showing the construct validity assessment for the entire scale, as well as the sub-scales.}
\label{tab:groups}
\end{table*}

\subsection{Discriminant Validity}
Next, we investigated the discriminant validity of our scale, i.e. determining that the scale was not a derivative of another concept. In our case, we wanted to verify that the TSRI was not a reflection of the users' personal qualities which lead to reflection. To that end, we compared the TSRI with the The Self-Reflection and Insight Scale~\cite{Grant2002TheSelf-consciousness}---a validated instrument for measuring self-regulatory processes related to reflection. We conducted a Pearson Product-Moment correlation test to investigate if the TSRI was correlated with the SRIS. We obtained a result of $r(54)=0.38$, $p<0.01$, indicating a weak to moderate correlation. This indicates that the TSRI is conceptually different, but the personality factors measured in SRIS may play a role in its scoring, likely as moderator variables. However, this results also confirms that SRIS can be used effectively to compare interactive technologies, reasonably irrespective of user personalities.
 
\subsection{Test-Retest Reliability}
Finally, we checked if \scaleshort{} could produce results that were reliable at different points in time, i.e. we tested its temporal stability. We conducted a third online survey for this purpose. In this survey, participants were asked to rate their tracking technology using the \scaleshort{}. The full survey and results for the test-retest are available in the auxiliary material. 

\subsubsection{Participants}
We administered the \scaleshort{} to a group of $n=20$ participants, aged $M=29.3$, $SD=13.1$, 10 female and 10 male, twice, with a 14-day break in between the studies. Participants were recruited through snowball sampling. The reported sample does not include participants who did not respond to our invitation to complete the survey for the second time. In line with the first survey, we indicated that participants needed to use tracking technology (e.g. smartwatch, tracking app) to partake in the survey. This ensured that participants used some form of an interactive technology that could facilitate reflection and could form informed opinions. 

\subsubsection{Results}
There is no consensus over what metrics to use to test for test-retest reliability~\cite{Boateng2018BestPF}. We used one of Boateng et al.'s suggestions and calculated the intra class correlation coefficient for fixed raters. The result was $\kappa=0.73$, $p<.01$. The 95\% confidence interval ranged from $0.42$ to $0.88$, According to Koo and Mae al.~\cite{koo_guideline_2016}, this indicates moderate reliability with a confidence interval between poor and excellent.

\section{Discussion}
In this section, we explain how the \scaleshort{} can be used in practice and we discuss the limitations of our approach and possibilities for further development.

\subsection{Operationalising Reflection}
In the tradition of scale development and psychological assessment, content validity is of primary importance. Thus, we followed a structured procedure in our scale development. Multiple psychological assessment instruments were developed, successfully used, and revised using this and similar processes, when theories about the construct (i.e. personal reflection) evolve and new data is acquired (e.g. with the means of our scale)~\cite{haynes1995content}. Our scale, therefore, offers an implicit definition of reflection. This constitutes a possible empirical operationalisation for the HCI field in the face of the lack of an analytically derived definition. Yet, we recognise that there is a need for further understanding and defining technology-supported reflection.

\subsection{Using the TSRI}
The \scaleshort{} can be used in the early stages of designing interactive systems for reflection. Because the scale is relatively short, it can be used for rapid feedback. Using the scale enables the evaluation and comparison of different designs for reflection. However, seeing that a multitude of different definitions of reflection are being used, this might mean that people have contrasting views as to what reflection entails. Thus, it is desirable to compare these different technologies within the same context. By evaluating systems in the same context, the effects of people having different understandings of the term reflection can be minimised. This implies that the scale is not suited to scoring reflection in absolute terms or reflection as an attitude. The \scaleshort{} is intended as an artefact-centric instrument that enables researchers and designers to compare different designs relatively from one another.

\subsubsection{Scoring and analysis}
The \scaleshort{} is scored on a seven-point Likert scale from Strongly Disagree (1) to Strongly Agree (7). The items of the scale are balanced over the dimensions and there are no reversed items, which makes the analysis of the scores a straightforward process. The sum of the scores indicates the extent to which an interactive system facilitates reflection. The lowest score on the scale is 9 and the highest is 63. Higher scores indicate that the artefact offers more support for reflection. The scores offer a transparent and actionable means for designers and researchers to use the \scaleshort{}.

\subsection{Limitations}
There are several limitations to the development of \scaleshort{}. In recruiting participants for the study we indicated on MTurk that participants needed to use tracking technology (i.e. smartwatch, tracking app) to partake in the survey. We chose to do that to ensure that participants used some form of an interactive technology that can facilitate reflection. Even though participants used a variety of technologies, with $46\%$ using their smartphones as tracking devices, our considerations were focused on understanding the reflection over self-generated data, collected over time through one's own performance. This might have influenced the formation of the scale, possibly making the scale a better fit particularly for personal informatics systems. We believe that developing related scales, which feature a less data-driven approach is a promising way to understand a wider range of systems which possibly support reflection. Finally, our user samples consisted primarily of Western European and US-based participants. We recognise that the scale, therefore, is constrained within the Western culture and might be prone to cultural biases. 

\section{Conclusion}
In this paper, we presented the development and evaluation of the \scale{} (\scaleshort{}). Through a review of the literature, we identified relevant concepts for reflection, which formed the basis for generating scale items. We then built a nine-item scale and \scaleshort{}'s validity and reliability. We demonstrated the scale's discriminant validity, its ability to differentiate between known groups and test-retest reliability. The \scaleshort{} enables researchers and designers to evaluate interactive systems designed for facilitating reflection. The scale can be used to rapidly compare prototypes or assess novel artefacts. 

\begin{acks}
This research was financially supported by Utrecht University's Focus Area: Sports and Society. We acknowledge the support of the Leibniz ScienceCampus Bremen Digital Public Health (lsc-diph.de), which is jointly funded by the Leibniz Association (W4/2018), the Federal State of Bremen and the Leibniz Institute for Prevention Research and Epidemiology---BIPS. 
\end{acks}

\bibliographystyle{ACM-Reference-Format}
\bibliography{references,acmart}


\begin{thebibliography}{31}


\ifx \showCODEN    \undefined \def \showCODEN     #1{\unskip}     \fi
\ifx \showDOI      \undefined \def \showDOI       #1{#1}\fi
\ifx \showISBNx    \undefined \def \showISBNx     #1{\unskip}     \fi
\ifx \showISBNxiii \undefined \def \showISBNxiii  #1{\unskip}     \fi
\ifx \showISSN     \undefined \def \showISSN      #1{\unskip}     \fi
\ifx \showLCCN     \undefined \def \showLCCN      #1{\unskip}     \fi
\ifx \shownote     \undefined \def \shownote      #1{#1}          \fi
\ifx \showarticletitle \undefined \def \showarticletitle #1{#1}   \fi
\ifx \showURL      \undefined \def \showURL       {\relax}        \fi
\providecommand\bibfield[2]{#2}
\providecommand\bibinfo[2]{#2}
\providecommand\natexlab[1]{#1}
\providecommand\showeprint[2][]{arXiv:#2}

\bibitem[\protect\citeauthoryear{Arakawa and Yakura}{Arakawa and
  Yakura}{2020}]%
        {Arakawa2020INWARD:Coaching}
\bibfield{author}{\bibinfo{person}{Riku Arakawa} {and} \bibinfo{person}{Hiromu
  Yakura}.} \bibinfo{year}{2020}\natexlab{}.
\newblock \showarticletitle{{INWARD: A Computer-Supported Tool for
  Video-Reflection Improves Efficiency and Effectiveness in Executive
  Coaching}}.
\newblock  (\bibinfo{year}{2020}), \bibinfo{pages}{1--13}.
\newblock
\showISBNx{9781450367080}
\urldef\tempurl%
\url{https://doi.org/10.1145/3313831.3376703}
\showDOI{\tempurl}


\bibitem[\protect\citeauthoryear{Aukes, Geertsma, Cohen-Schotanus, Zwierstra,
  and Slaets}{Aukes et~al\mbox{.}}{2007}]%
        {Aukes2007TheEducation}
\bibfield{author}{\bibinfo{person}{Leo~C. Aukes}, \bibinfo{person}{Jelle
  Geertsma}, \bibinfo{person}{Janke Cohen-Schotanus}, \bibinfo{person}{Rein~P.
  Zwierstra}, {and} \bibinfo{person}{Joris~P.J. Slaets}.}
  \bibinfo{year}{2007}\natexlab{}.
\newblock \showarticletitle{{The development of a scale to measure personal
  reflection in medical practice and education}}.
\newblock \bibinfo{journal}{\emph{Medical Teacher}} \bibinfo{volume}{29},
  \bibinfo{number}{2-3} (\bibinfo{year}{2007}), \bibinfo{pages}{177--182}.
\newblock
\showISSN{0142159X}
\urldef\tempurl%
\url{https://doi.org/10.1080/01421590701299272}
\showDOI{\tempurl}


\bibitem[\protect\citeauthoryear{Ayobi, Marshall, and Cox}{Ayobi
  et~al\mbox{.}}{2020}]%
        {Ayobi2020TracklySelf-Care}
\bibfield{author}{\bibinfo{person}{Amid Ayobi}, \bibinfo{person}{Paul
  Marshall}, {and} \bibinfo{person}{Anna~L Cox}.}
  \bibinfo{year}{2020}\natexlab{}.
\newblock \showarticletitle{{Trackly : A Customisable and Pictorial
  Self-Tracking App to Support Agency in Multiple Sclerosis Self-Care}}.
\newblock \bibinfo{journal}{\emph{Proceedings of the SIGCHI Conference on Human
  Factors in Computing Systems}} (\bibinfo{year}{2020}),
  \bibinfo{pages}{1--15}.
\newblock
\showISBNx{9781450367080}


\bibitem[\protect\citeauthoryear{Baumer, Katz, Freeman, Adams, Gonzales,
  Pollak, Retelny, Niederdeppe, Olson, and Gay}{Baumer et~al\mbox{.}}{2012}]%
        {Baumer2012PrescriptiveHealth}
\bibfield{author}{\bibinfo{person}{Eric~P.S. Baumer},
  \bibinfo{person}{Sherri~Jean Katz}, \bibinfo{person}{Jill~E. Freeman},
  \bibinfo{person}{Phil Adams}, \bibinfo{person}{Amy~L. Gonzales},
  \bibinfo{person}{John Pollak}, \bibinfo{person}{Daniela Retelny},
  \bibinfo{person}{Jeff Niederdeppe}, \bibinfo{person}{Christine~M. Olson},
  {and} \bibinfo{person}{Geri~K. Gay}.} \bibinfo{year}{2012}\natexlab{}.
\newblock \showarticletitle{{Prescriptive persuasion and open-ended social
  awareness: Expanding the design space of mobile health}}.
\newblock \bibinfo{journal}{\emph{Proceedings of the ACM Conference on Computer
  Supported Cooperative Work, CSCW}} (\bibinfo{year}{2012}),
  \bibinfo{pages}{475--484}.
\newblock
\showISBNx{9781450310864}
\urldef\tempurl%
\url{https://doi.org/10.1145/2145204.2145279}
\showDOI{\tempurl}


\bibitem[\protect\citeauthoryear{Baumer, Khovanskaya, Matthews, Reynolds,
  Sosik, and Gay}{Baumer et~al\mbox{.}}{2014}]%
        {Baumer2014ReviewingDesign}
\bibfield{author}{\bibinfo{person}{Eric~P.S. Baumer}, \bibinfo{person}{Vera
  Khovanskaya}, \bibinfo{person}{Mark Matthews}, \bibinfo{person}{Lindsay
  Reynolds}, \bibinfo{person}{Victoria~Schwanda Sosik}, {and}
  \bibinfo{person}{Geri Gay}.} \bibinfo{year}{2014}\natexlab{}.
\newblock \showarticletitle{{Reviewing reflection: On the use of reflection in
  interactive system design}}.
\newblock \bibinfo{journal}{\emph{Proceedings of the Conference on Designing
  Interactive Systems: Processes, Practices, Methods, and Techniques, DIS}}
  (\bibinfo{year}{2014}), \bibinfo{pages}{93--102}.
\newblock
\showISBNx{9781450329026}
\urldef\tempurl%
\url{https://doi.org/10.1145/2598510.2598598}
\showDOI{\tempurl}


\bibitem[\protect\citeauthoryear{Baumer}{Baumer}{2015}]%
        {Baumer_reflectiveinformatics}
\bibfield{author}{\bibinfo{person}{Eric P~S Baumer}.}
  \bibinfo{year}{2015}\natexlab{}.
\newblock \showarticletitle{{Reflective Informatics: Conceptual Dimensions for
  Designing Technologies of Reflection}}. In
  \bibinfo{booktitle}{\emph{Proceedings of the 33rd Annual ACM Conference on
  Human Factors in Computing Systems}} \emph{(\bibinfo{series}{CHI ’15})}.
  \bibinfo{publisher}{Association for Computing Machinery},
  \bibinfo{address}{New York, NY, USA}, \bibinfo{pages}{585–594}.
\newblock
\showISBNx{9781450331456}
\urldef\tempurl%
\url{https://doi.org/10.1145/2702123.2702234}
\showDOI{\tempurl}


\bibitem[\protect\citeauthoryear{Boateng, Neilands, Frongillo,
  Melgar-Qui{\~{n}}onez, and Young}{Boateng et~al\mbox{.}}{2018}]%
        {Boateng2018BestPF}
\bibfield{author}{\bibinfo{person}{Godfred~O. Boateng},
  \bibinfo{person}{Torsten~B. Neilands}, \bibinfo{person}{Edward~A. Frongillo},
  \bibinfo{person}{Hugo~R. Melgar-Qui{\~{n}}onez}, {and}
  \bibinfo{person}{Sera~L. Young}.} \bibinfo{year}{2018}\natexlab{}.
\newblock \showarticletitle{{Best Practices for Developing and Validating
  Scales for Health, Social, and Behavioral Research: A Primer}}.
\newblock \bibinfo{journal}{\emph{Frontiers in Public Health}}
  \bibinfo{volume}{6} (\bibinfo{year}{2018}).
\newblock
\showISSN{2296-2565}
\urldef\tempurl%
\url{https://doi.org/10.3389/fpubh.2018.00149}
\showDOI{\tempurl}


\bibitem[\protect\citeauthoryear{Clayphan, Martinez-Maldonado, and
  Kay}{Clayphan et~al\mbox{.}}{2017}]%
        {Clayphan2017ASpace}
\bibfield{author}{\bibinfo{person}{Andrew~John Clayphan},
  \bibinfo{person}{Roberto Martinez-Maldonado}, {and} \bibinfo{person}{Judy
  Kay}.} \bibinfo{year}{2017}\natexlab{}.
\newblock \showarticletitle{{A student-facing dashboard for supporting
  sensemaking about the brainstorm process at a multi-surface space}}.
\newblock \bibinfo{journal}{\emph{ACM International Conference Proceeding
  Series}} (\bibinfo{year}{2017}), \bibinfo{pages}{49--58}.
\newblock
\showISBNx{9781450353793}
\urldef\tempurl%
\url{https://doi.org/10.1145/3152771.3152777}
\showDOI{\tempurl}


\bibitem[\protect\citeauthoryear{Epstein, Ping, Fogarty, and Munson}{Epstein
  et~al\mbox{.}}{2015}]%
        {Epstein2015AInformatics}
\bibfield{author}{\bibinfo{person}{Daniel~A. Epstein}, \bibinfo{person}{An
  Ping}, \bibinfo{person}{James Fogarty}, {and} \bibinfo{person}{Sean~A.
  Munson}.} \bibinfo{year}{2015}\natexlab{}.
\newblock \showarticletitle{{A lived informatics model of personal
  informatics}}.
\newblock \bibinfo{journal}{\emph{UbiComp 2015 - Proceedings of the 2015 ACM
  International Joint Conference on Pervasive and Ubiquitous Computing}}
  (\bibinfo{year}{2015}), \bibinfo{pages}{731--742}.
\newblock
\showISBNx{9781450335744}
\urldef\tempurl%
\url{https://doi.org/10.1145/2750858.2804250}
\showDOI{\tempurl}


\bibitem[\protect\citeauthoryear{Fleck and Fitzpatrick}{Fleck and
  Fitzpatrick}{2010}]%
        {Fleck2010ReflectingLandscape}
\bibfield{author}{\bibinfo{person}{Rowanne Fleck} {and}
  \bibinfo{person}{Geraldine Fitzpatrick}.} \bibinfo{year}{2010}\natexlab{}.
\newblock \showarticletitle{{Reflecting on reflection: Framing a design
  landscape}}.
\newblock \bibinfo{journal}{\emph{ACM International Conference Proceeding
  Series}} (\bibinfo{year}{2010}), \bibinfo{pages}{216--223}.
\newblock
\showISBNx{9781450305020}
\urldef\tempurl%
\url{https://doi.org/10.1145/1952222.1952269}
\showDOI{\tempurl}


\bibitem[\protect\citeauthoryear{Grant, Franklin, and Langford}{Grant
  et~al\mbox{.}}{2002}]%
        {Grant2002TheSelf-consciousness}
\bibfield{author}{\bibinfo{person}{Anthony~M Grant}, \bibinfo{person}{John
  Franklin}, {and} \bibinfo{person}{Peter Langford}.}
  \bibinfo{year}{2002}\natexlab{}.
\newblock \showarticletitle{{The Self-Reflection and Insight Scale: A new
  measure of private self-consciousness}}.
\newblock \bibinfo{journal}{\emph{Social Behavior and Personality}}
  \bibinfo{volume}{30}, \bibinfo{number}{8} (\bibinfo{year}{2002}),
  \bibinfo{pages}{821--836}.
\newblock
\urldef\tempurl%
\url{https://doi.org/10.2224/sbp.2002.30.8.821}
\showDOI{\tempurl}


\bibitem[\protect\citeauthoryear{Haynes, Richard, and Kubany}{Haynes
  et~al\mbox{.}}{1995}]%
        {haynes1995content}
\bibfield{author}{\bibinfo{person}{Stephen~N Haynes}, \bibinfo{person}{David
  Richard}, {and} \bibinfo{person}{Edward~S Kubany}.}
  \bibinfo{year}{1995}\natexlab{}.
\newblock \showarticletitle{Content validity in psychological assessment: A
  functional approach to concepts and methods.}
\newblock \bibinfo{journal}{\emph{Psychological assessment}}
  \bibinfo{volume}{7}, \bibinfo{number}{3} (\bibinfo{year}{1995}),
  \bibinfo{pages}{238}.
\newblock


\bibitem[\protect\citeauthoryear{Isaacs, Konrad, Walendowski, Lennig, Hollis,
  and Whittaker}{Isaacs et~al\mbox{.}}{2013}]%
        {Isaacs2013EchoesWell-being}
\bibfield{author}{\bibinfo{person}{Ellen Isaacs}, \bibinfo{person}{Artie
  Konrad}, \bibinfo{person}{Alan Walendowski}, \bibinfo{person}{Thomas Lennig},
  \bibinfo{person}{Victoria Hollis}, {and} \bibinfo{person}{Steve Whittaker}.}
  \bibinfo{year}{2013}\natexlab{}.
\newblock \showarticletitle{{Echoes from the past: How technology mediated
  reflection improves well-being}}.
\newblock \bibinfo{journal}{\emph{Conference on Human Factors in Computing
  Systems - Proceedings}} (\bibinfo{year}{2013}), \bibinfo{pages}{1071--1080}.
\newblock
\showISBNx{9781450318990}
\urldef\tempurl%
\url{https://doi.org/10.1145/2470654.2466137}
\showDOI{\tempurl}


\bibitem[\protect\citeauthoryear{Koo and Li}{Koo and Li}{2016}]%
        {koo_guideline_2016}
\bibfield{author}{\bibinfo{person}{Terry~K. Koo} {and} \bibinfo{person}{Mae~Y.
  Li}.} \bibinfo{year}{2016}\natexlab{}.
\newblock \showarticletitle{A {Guideline} of {Selecting} and {Reporting}
  {Intraclass} {Correlation} {Coefficients} for {Reliability} {Research}}.
\newblock \bibinfo{journal}{\emph{Journal of Chiropractic Medicine}}
  \bibinfo{volume}{15}, \bibinfo{number}{2} (\bibinfo{year}{2016}),
  \bibinfo{pages}{155 -- 163}.
\newblock
\showISSN{1556-3707}
\urldef\tempurl%
\url{https://doi.org/10.1016/j.jcm.2016.02.012}
\showDOI{\tempurl}


\bibitem[\protect\citeauthoryear{Li, Dey, and Forlizzi}{Li
  et~al\mbox{.}}{2010}]%
        {Li_2010_new}
\bibfield{author}{\bibinfo{person}{Ian Li}, \bibinfo{person}{Anind Dey}, {and}
  \bibinfo{person}{Jodi Forlizzi}.} \bibinfo{year}{2010}\natexlab{}.
\newblock \showarticletitle{{A Stage-Based Model of Personal Informatics
  Systems}}. In \bibinfo{booktitle}{\emph{Proceedings of the SIGCHI Conference
  on Human Factors in Computing Systems}} \emph{(\bibinfo{series}{CHI ’10})}.
  \bibinfo{publisher}{Association for Computing Machinery},
  \bibinfo{address}{New York, NY, USA}, \bibinfo{pages}{557–566}.
\newblock
\showISBNx{9781605589299}
\urldef\tempurl%
\url{https://doi.org/10.1145/1753326.1753409}
\showDOI{\tempurl}


\bibitem[\protect\citeauthoryear{Marsden}{Marsden}{2013}]%
        {marsden_2013}
\bibfield{author}{\bibinfo{person}{Nicola Marsden}.}
  \bibinfo{year}{2013}\natexlab{}.
\newblock \showarticletitle{{Attitudes towards Online Communication: An
  Exploratory Factor Analysis}}. In \bibinfo{booktitle}{\emph{Proceedings of
  the 2013 Annual Conference on Computers and People Research}}
  \emph{(\bibinfo{series}{SIGMIS-CPR '13})}. \bibinfo{publisher}{Association
  for Computing Machinery}, \bibinfo{address}{New York, NY, USA},
  \bibinfo{pages}{147–152}.
\newblock
\showISBNx{9781450319751}
\urldef\tempurl%
\url{https://doi.org/10.1145/2487294.2487326}
\showDOI{\tempurl}


\bibitem[\protect\citeauthoryear{Mejia and Yarosh}{Mejia and Yarosh}{2017}]%
        {Mejia_Yarosh_2017}
\bibfield{author}{\bibinfo{person}{Kenya Mejia} {and} \bibinfo{person}{Svetlana
  Yarosh}.} \bibinfo{year}{2017}\natexlab{}.
\newblock \showarticletitle{{A Nine-Item Questionnaire for Measuring the Social
  Disfordance of Mediated Social Touch Technologies}}.
\newblock \bibinfo{journal}{\emph{Proc. ACM Hum.-Comput. Interact.}}
  \bibinfo{volume}{1}, \bibinfo{number}{CSCW} (\bibinfo{date}{12}
  \bibinfo{year}{2017}).
\newblock
\urldef\tempurl%
\url{https://doi.org/10.1145/3134712}
\showDOI{\tempurl}


\bibitem[\protect\citeauthoryear{Mols, Van Den~Hoven, and Eggen}{Mols
  et~al\mbox{.}}{2016}]%
        {Mols2016InformingPractices}
\bibfield{author}{\bibinfo{person}{Ine Mols}, \bibinfo{person}{Elise Van
  Den~Hoven}, {and} \bibinfo{person}{Berry Eggen}.}
  \bibinfo{year}{2016}\natexlab{}.
\newblock \showarticletitle{{Informing design for reflection: An overview of
  current everyday practices}}.
\newblock \bibinfo{journal}{\emph{ACM International Conference Proceeding
  Series}}  \bibinfo{volume}{23-27-Octo} (\bibinfo{year}{2016}).
\newblock
\showISBNx{9781450347631}
\urldef\tempurl%
\url{https://doi.org/10.1145/2971485.2971494}
\showDOI{\tempurl}


\bibitem[\protect\citeauthoryear{Mols, Van Den~Hoven, and Eggen}{Mols
  et~al\mbox{.}}{2020}]%
        {Mols2020EverydayDott}
\bibfield{author}{\bibinfo{person}{Ine Mols}, \bibinfo{person}{Elise Van
  Den~Hoven}, {and} \bibinfo{person}{Berry Eggen}.}
  \bibinfo{year}{2020}\natexlab{}.
\newblock \showarticletitle{{Everyday life reflection: Exploring media
  interaction with balance, cogito {\&} dott}}.
\newblock \bibinfo{journal}{\emph{TEI 2020 - Proceedings of the 14th
  International Conference on Tangible, Embedded, and Embodied Interaction}}
  (\bibinfo{year}{2020}), \bibinfo{pages}{67--79}.
\newblock
\showISBNx{9781450361071}
\urldef\tempurl%
\url{https://doi.org/10.1145/3374920.3374928}
\showDOI{\tempurl}


\bibitem[\protect\citeauthoryear{Moon}{Moon}{1999}]%
        {book}
\bibfield{author}{\bibinfo{person}{Jennifer Moon}.}
  \bibinfo{year}{1999}\natexlab{}.
\newblock \bibinfo{booktitle}{\emph{{Reflection in learning {\&} professional
  development}}}.
\newblock \bibinfo{publisher}{RoutledgeFalmer}, \bibinfo{address}{New York, NY,
  USA}. 229 pages.
\newblock
\showISBNx{9780749434526}
\urldef\tempurl%
\url{https://doi.org/10.4324/9780203822296}
\showDOI{\tempurl}


\bibitem[\protect\citeauthoryear{Niess and Wo{\'{z}}niak}{Niess and
  Wo{\'{z}}niak}{2018}]%
        {Niess2018SupportingModel}
\bibfield{author}{\bibinfo{person}{Jasmin Niess} {and}
  \bibinfo{person}{Paweł~W. Wo{\'{z}}niak}.} \bibinfo{year}{2018}\natexlab{}.
\newblock \showarticletitle{{Supporting meaningful personal fitness: The
  tracker goal Evolution Model}}.
\newblock \bibinfo{journal}{\emph{Conference on Human Factors in Computing
  Systems - Proceedings}}  \bibinfo{volume}{2018-April} (\bibinfo{year}{2018}),
  \bibinfo{pages}{1--12}.
\newblock
\showISBNx{9781450356206}
\urldef\tempurl%
\url{https://doi.org/10.1145/3173574.3173745}
\showDOI{\tempurl}


\bibitem[\protect\citeauthoryear{Ortiz and Fox~Harrell}{Ortiz and
  Fox~Harrell}{2018}]%
        {Ortiz2018EnablingChimeria:Grayscale}
\bibfield{author}{\bibinfo{person}{Pablo Ortiz} {and} \bibinfo{person}{D.
  Fox~Harrell}.} \bibinfo{year}{2018}\natexlab{}.
\newblock \showarticletitle{{Enabling critical self-reflection through roleplay
  with chimeria:Grayscale}}.
\newblock \bibinfo{journal}{\emph{CHI PLAY 2018 - Proceedings of the 2018
  Annual Symposium on Computer-Human Interaction in Play}}
  (\bibinfo{year}{2018}), \bibinfo{pages}{365--380}.
\newblock
\showISBNx{9781450356244}
\urldef\tempurl%
\url{https://doi.org/10.1145/3242671.3242687}
\showDOI{\tempurl}


\bibitem[\protect\citeauthoryear{Rivera-Pelayo, Fessl, M{\"{u}}ller, and
  Pammer}{Rivera-Pelayo et~al\mbox{.}}{2017}]%
        {Rivera-Pelayo2017IntroducingWork}
\bibfield{author}{\bibinfo{person}{Verónica Rivera-Pelayo},
  \bibinfo{person}{Angela Fessl}, \bibinfo{person}{Lars M{\"{u}}ller}, {and}
  \bibinfo{person}{Viktoria Pammer}.} \bibinfo{year}{2017}\natexlab{}.
\newblock \showarticletitle{{Introducing Mood Self-Tracking at Work}}.
\newblock \bibinfo{journal}{\emph{ACM Transactions on Computer-Human
  Interaction}} \bibinfo{volume}{24}, \bibinfo{number}{1}
  (\bibinfo{year}{2017}), \bibinfo{pages}{1--28}.
\newblock
\showISSN{1073-0516}
\urldef\tempurl%
\url{https://doi.org/10.1145/3014058}
\showDOI{\tempurl}


\bibitem[\protect\citeauthoryear{Sch{\"{o}}n}{Sch{\"{o}}n}{1983}]%
        {Schon1983ThePractitioner}
\bibfield{author}{\bibinfo{person}{Donald~A. Sch{\"{o}}n}.}
  \bibinfo{year}{1983}\natexlab{}.
\newblock \bibinfo{booktitle}{\emph{{The Reﬂective Practitioner}}}.
\newblock
\showISBNx{046506874X}
\showISSN{04650687}
\urldef\tempurl%
\url{https://doi.org/10.1542/peds.2005-0209}
\showDOI{\tempurl}


\bibitem[\protect\citeauthoryear{Slovak, Frauenberger, and Fitzpatrick}{Slovak
  et~al\mbox{.}}{2017}]%
        {Slovak2017ReflectiveReflection}
\bibfield{author}{\bibinfo{person}{Petr Slovak}, \bibinfo{person}{Chris
  Frauenberger}, {and} \bibinfo{person}{Geraldine Fitzpatrick}.}
  \bibinfo{year}{2017}\natexlab{}.
\newblock \showarticletitle{{Reflective practicum: A framework of sensitising
  concepts to design for transformative reflection}}.
\newblock \bibinfo{journal}{\emph{Conference on Human Factors in Computing
  Systems - Proceedings}}  \bibinfo{volume}{2017-May} (\bibinfo{year}{2017}),
  \bibinfo{pages}{2696--2707}.
\newblock
\showISBNx{9781450346559}
\urldef\tempurl%
\url{https://doi.org/10.1145/3025453.3025516}
\showDOI{\tempurl}


\bibitem[\protect\citeauthoryear{Suh, Shahriaree, Hekler, and Kientz}{Suh
  et~al\mbox{.}}{2016}]%
        {Suh2016Developing}
\bibfield{author}{\bibinfo{person}{Hyewon Suh}, \bibinfo{person}{Nina
  Shahriaree}, \bibinfo{person}{Eric~B. Hekler}, {and}
  \bibinfo{person}{Julie~A. Kientz}.} \bibinfo{year}{2016}\natexlab{}.
\newblock \showarticletitle{Developing and Validating the User Burden Scale: A
  Tool for Assessing User Burden in Computing Systems}. In
  \bibinfo{booktitle}{\emph{Proceedings of the 2016 CHI Conference on Human
  Factors in Computing Systems}} (San Jose, California, USA)
  \emph{(\bibinfo{series}{CHI '16})}. \bibinfo{publisher}{Association for
  Computing Machinery}, \bibinfo{address}{New York, NY, USA},
  \bibinfo{pages}{3988–3999}.
\newblock
\showISBNx{9781450333627}
\urldef\tempurl%
\url{https://doi.org/10.1145/2858036.2858448}
\showDOI{\tempurl}


\bibitem[\protect\citeauthoryear{Thieme, Comber, Miebach, Weeden, Kr{\"{a}}mer,
  Lawson, and Olivier}{Thieme et~al\mbox{.}}{2012}]%
        {Thieme2012weveLifestyles}
\bibfield{author}{\bibinfo{person}{Anja Thieme}, \bibinfo{person}{Rob Comber},
  \bibinfo{person}{Julia Miebach}, \bibinfo{person}{Jack Weeden},
  \bibinfo{person}{Nicole Kr{\"{a}}mer}, \bibinfo{person}{Shaun Lawson}, {and}
  \bibinfo{person}{Patrick Olivier}.} \bibinfo{year}{2012}\natexlab{}.
\newblock \showarticletitle{{"we've bin watching you" - Designing for
  reflection and social persuasion to promote sustainable lifestyles}}.
\newblock \bibinfo{journal}{\emph{Conference on Human Factors in Computing
  Systems - Proceedings}} (\bibinfo{year}{2012}), \bibinfo{pages}{2337--2346}.
\newblock
\showISBNx{9781450310154}
\urldef\tempurl%
\url{https://doi.org/10.1145/2207676.2208394}
\showDOI{\tempurl}


\bibitem[\protect\citeauthoryear{Tondello, Wehbe, Diamond, Busch, Marczewski,
  and Nacke}{Tondello et~al\mbox{.}}{2016}]%
        {Tondello_2016}
\bibfield{author}{\bibinfo{person}{Gustavo~F Tondello}, \bibinfo{person}{Rina~R
  Wehbe}, \bibinfo{person}{Lisa Diamond}, \bibinfo{person}{Marc Busch},
  \bibinfo{person}{Andrzej Marczewski}, {and} \bibinfo{person}{Lennart~E
  Nacke}.} \bibinfo{year}{2016}\natexlab{}.
\newblock \showarticletitle{{The Gamification User Types Hexad Scale}}. In
  \bibinfo{booktitle}{\emph{Proceedings of the 2016 Annual Symposium on
  Computer-Human Interaction in Play}} \emph{(\bibinfo{series}{CHI PLAY '16})}.
  \bibinfo{publisher}{Association for Computing Machinery},
  \bibinfo{address}{New York, NY, USA}, \bibinfo{pages}{229–243}.
\newblock
\showISBNx{9781450344562}
\urldef\tempurl%
\url{https://doi.org/10.1145/2967934.2968082}
\showDOI{\tempurl}


\bibitem[\protect\citeauthoryear{Valkanova, Jorda, Tomitsch, and
  Vande~Moere}{Valkanova et~al\mbox{.}}{2013}]%
        {Valkanova_2013_new}
\bibfield{author}{\bibinfo{person}{Nina Valkanova}, \bibinfo{person}{Sergi
  Jorda}, \bibinfo{person}{Martin Tomitsch}, {and} \bibinfo{person}{Andrew
  Vande~Moere}.} \bibinfo{year}{2013}\natexlab{}.
\newblock \showarticletitle{{Reveal-It! The Impact of a Social Visualization
  Projection on Public Awareness and Discourse}}. In
  \bibinfo{booktitle}{\emph{Proceedings of the SIGCHI Conference on Human
  Factors in Computing Systems}} \emph{(\bibinfo{series}{CHI ’13})}.
  \bibinfo{publisher}{Association for Computing Machinery},
  \bibinfo{address}{New York, NY, USA}, \bibinfo{pages}{3461–3470}.
\newblock
\showISBNx{9781450318990}
\urldef\tempurl%
\url{https://doi.org/10.1145/2470654.2466476}
\showDOI{\tempurl}


\bibitem[\protect\citeauthoryear{Webb, Linder, Kerne, Lupfer, Qu, Poffenberger,
  and Revia}{Webb et~al\mbox{.}}{2013}]%
        {Webb_2013_new}
\bibfield{author}{\bibinfo{person}{Andrew~M Webb}, \bibinfo{person}{Rhema
  Linder}, \bibinfo{person}{Andruid Kerne}, \bibinfo{person}{Nic Lupfer},
  \bibinfo{person}{Yin Qu}, \bibinfo{person}{Bryant Poffenberger}, {and}
  \bibinfo{person}{Colton Revia}.} \bibinfo{year}{2013}\natexlab{}.
\newblock \showarticletitle{{Promoting Reflection and Interpretation in
  Education: Curating Rich Bookmarks as Information Composition}}. In
  \bibinfo{booktitle}{\emph{Proceedings of the 9th ACM Conference on Creativity
  {\&}amp; Cognition}} \emph{(\bibinfo{series}{C{\&}amp;C ’13})}.
  \bibinfo{publisher}{Association for Computing Machinery},
  \bibinfo{address}{New York, NY, USA}, \bibinfo{pages}{53–62}.
\newblock
\showISBNx{9781450321501}
\urldef\tempurl%
\url{https://doi.org/10.1145/2466627.2466636}
\showDOI{\tempurl}


\bibitem[\protect\citeauthoryear{Yarosh, Markopoulos, and Abowd}{Yarosh
  et~al\mbox{.}}{2014}]%
        {Yarosh2014CSCW}
\bibfield{author}{\bibinfo{person}{Svetlana Yarosh}, \bibinfo{person}{Panos
  Markopoulos}, {and} \bibinfo{person}{Gregory~D. Abowd}.}
  \bibinfo{year}{2014}\natexlab{}.
\newblock \showarticletitle{Towards a Questionnaire for Measuring Affective
  Benefits and Costs of Communication Technologies}. In
  \bibinfo{booktitle}{\emph{Proceedings of the 17th ACM Conference on Computer
  Supported Cooperative Work \& Social Computing}} (Baltimore, Maryland, USA)
  \emph{(\bibinfo{series}{CSCW '14})}. \bibinfo{publisher}{Association for
  Computing Machinery}, \bibinfo{address}{New York, NY, USA},
  \bibinfo{pages}{84–96}.
\newblock
\showISBNx{9781450325400}
\urldef\tempurl%
\url{https://doi.org/10.1145/2531602.2531634}
\showDOI{\tempurl}


\end{thebibliography}
\end{document}